# The True (?) Story of Hilbert's Infinite Hotel

Helge Kragh[*]

**Abstract:** What is known as "Hilbert's hotel" is a story of an imaginary hotel with infinitely many rooms that illustrates the bizarre consequences of assuming an actual infinity of objects or events. Since the 1970s it has been used in a variety of arguments, some of them relating to cosmology and others to philosophy and theology. For a long time it has remained unknown whether David Hilbert actually proposed the thought experiment named after him, or whether it was merely a piece of mathematical folklore. It turns out that Hilbert introduced his hotel in a lecture of January 1924, but without publishing it. The counter-intuitive hotel only became better known in 1947, when George Gamow described it in a book, and it took nearly three more decades until it attracted wide interest in scientific, philosophical, and theological contexts. The paper outlines the origin and early history of Hilbert's hotel paradox. At the same time it retracts the author's earlier conclusion that the paradox was originally due to Gamow.

**1. Introduction**

On a dark desert highway a tired driver passes one more hotel with a "No Vacancy" sign. But this time the hotel looks exceedingly large and so he goes in to see if there might nonetheless be a room for him:

> The clerk said, "No problem. Here's what can be done –
> We'll move those in a room to the next higher one.
> That will free up the first room and that's where you can stay."
> I tried understanding that as I heard him say:

> CHORUS: "Welcome to the HOTEL called INFINITY –
> Where every room is full (every room is full)
> Yet there's room for more.
> Yeah, plenty of room at the HOTEL called INFINITY –
> Move 'em down the floor (move 'em down the floor)
> To make room for more."

---

[*] Centre for Science Studies, Department of Physics and Astronomy, Aarhus University, 8000 Aarhus, Denmark. E-mail: helge.kragh@ivs.au.dk.

What the tired driver experienced was the famous "Hilbert's hotel," a thought experiment illustrating the paradoxical nature of actual infinities. As new guests check in he constantly has to move to rooms with higher numbers, causing him a sleepless night ("Never more will I confuse a Hilton with a Hilbert Hotel!"). But Hilbert's hotel has an advantage that Hilton's cannot match:

> Last thing I remember at the end of my stay –
> It was time to pay the bill but I had no means to pay.
> The man in 19 smiled, "Your bill is on me.
> 20 pays mine, and so on, so you get yours for free!"[1]

The mathematical paradox about infinite sets associated with Hilbert's name envisages a hotel with a countable infinity of rooms, that is, rooms that can be placed in a one-to-one correspondence with the natural numbers. All rooms in the hotel are occupied. Now suppose that a new guest arrives – will it be possible to find a free room for him or her? Surprisingly, the answer is yes. He (or she) may be accommodated in room 1, while the guest in this room is moved to room 2, the guest in room 2 moves to room 3, and so on. Since there is no last room, the newcomer can be accommodated without any of the guests having to leave the hotel. Hilbert's remarkable hotel can even accommodate a countable infinity of new guests without anyone leaving it. The guests in rooms with the number $n$ only have to change to rooms $2n$, which will leave an infinite number of odd-numbered rooms available for the infinite number of new guests. What the parable tells us is that the statement "all rooms are occupied" does not imply that "there is no more space for new guests." This is strange indeed, although it is not, strictly speaking, a paradox in the logical sense of the term. Yet it is so counter-intuitive that it suggests that countable actual infinities do not belong to the real world we live in.

The highly counter-intuitive hotel is a standard story in discourses on mathematics, philosophy, cosmology, and theology, and yet the historical origin of Hilbert's famous hotel has rarely if ever been investigated. By far most sources assume, explicitly or implicitly, that it was constructed by the German mathematician David Hilbert, one of the giants of twentieth-century mathematics. After all, why else use the eponymous label "Hilbert's hotel"?

Mathematical writers Rudy Rucker and Eli Maor helped making Hilbert's hotel widely known in the 1980s by including it in their popular books on the mysteries of the infinite. According to Rucker [1982, p. 73], "The famous mathematician David Hilbert used to illustrate his popular lectures with stories about a hotel with infinitely many rooms." Maor [1987, p. vii] more cautiously referred to "a story attributed to David Hilbert." The informative Wikipedia article discusses the paradox as "presented by David Hilbert in the 1920s," and an equally informative article in *The New York Times* of 2010 claims to "follow an approach introduced by Hilbert himself … by telling a parable about a grand hotel" [Strogatz 2010]. None of the many writings on the infinite hotel provides any information of the source of Hilbert's hotel. However, Barrow [2005, p. 284] notes that Hilbert never wrote about his hotel and that it became better known only after George Gamow described it in a book of 1947.

---

[1] "Hotel Infinity," poem in six verses by Lawrence Mark Lesser, first published in *American Mathematical Monthly* **113** (2006): 704. See also Glaz 2011, p. 177.

Attempting to find an answer to the question of the origin of Hilbert's hotel, in a recent communication to *Arxiv* I suggested that the originator was most likely Gamow, "who came upon the idea independently of Hilbert and only after his death in 1943."[2] Since I could find no trace of it in Hilbert's publications, there was only "the remote possibility that Hilbert actually discussed the grand hotel in some unpublished lecture or informal talk," which I mistakenly thought was unlikely. As I was soon informed by Tilman Sauer, Hilbert introduced his story about the hotel in unpublished lectures in the winter semester 1924-1925. These lectures have only recently appeared in print [Hilbert 2013].

## 2. Hilbert, Cantor, and the infinite

The discussion of Hilbert's hotel relates to the old question of whether an actual, as opposed to a potential, infinity is possible. According to Georg Cantor's theory of transfinite numbers, dating from the 1880s, this is indeed the case, namely in the sense that the concept of the actual infinite is logically consistent and operationally useful [Dauben 1979; Cantor 1962]. But one thing is mathematical consistency, another and more crucial question is whether an actual infinite can be instantiated in the real world as examined by the physicists and astronomers. From Cantor's standpoint, which can be characterized as essentially Platonic, numbers and other mathematical constructs had a permanent existence and were as real as – nay, were

---

[2] The *Arxiv* paper has now been replaced by the present one. The case illustrates that hypotheses in history are falsifiable, just as they are in the sciences. The original hypothesis of Gamow as the founder of Hilbert's hotel was proved wrong by new and better data.

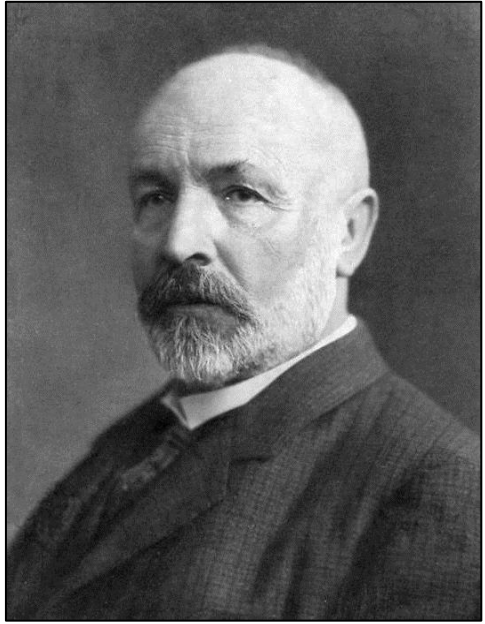
Fig. 1. Georg Cantor (1845-1918)

more real than – the ephemeral sense impressions on which the existence of physical objects and phenomena are based. From this position it was largely irrelevant whether or not the physical universe contains an infinite number of stars.

Contrary to some other contemporary mathematicians, including Leopold Kronecker and Henri Poincaré, Hilbert was greatly impressed by Cantor's set theory. This he made clear in a semi-popular lecture course he gave in Göttingen during the winter semester 1924-1925 and in which he dealt at length with the infinite in mathematics, physics, and astronomy. A few months later he repeated the message in a wide-ranging lecture on the infinite given in Münster on 4 June 1925. The occasion was a session organized by the Westphalian Mathematical Society to celebrate the mathematical work of Karl Weierstrass. Hilbert's Münster address drew extensively on his previous lecture course, except that it was more technical and omitted many examples. One of them was the infinite hotel. "No one shall expel us from the paradise which Cantor has created for us," Hilbert famously declared in his Münster address [Hilbert 1925, p. 170; Reid 1970, pp. 175-177; Benacerraf and Putnam 1983, pp. 134-151]. On the other hand, he did not believe that the actual infinities defined by Cantor had anything to do with the real world. This is what he said about the issue [Hilbert 1925, p. 190]:

> The infinity is nowhere to be found in reality. It neither exists in nature nor provides a legitimate basis for rational thought – a remarkable harmony between being and

> thought. … The role that remains for the infinite is solely that of an idea – if one means by an idea, in Kant's terminology, a concept of reason which transcends all experience and which completes the concrete as a totality – that of an idea which we may unhesitatingly trust within the framework erected by our [mathematical] theory.

From the point of view of an objective idealist like Cantor, Hilbert's exclusion of the infinite from reality was hardly a restriction at all. In spite of his defense of the reality of infinite numbers, he maintained that the physical universe was finite in both time and space. To claim otherwise, Cantor said in a letter of 1887 to Aloys von Schmid, a Munich professor of theology, was "monstrous nonsense" [Kragh 2008, p. 96; Cantor 1962, pp. 370-377].

Hilbert's addresses in Göttingen and Münster also merit attention because of their references to microphysics and cosmology. On the one hand, he said in Münster, the new atomic and quantum physics had demonstrated that the infinite divisibility of space and matter was nothing but an abstract idea inapplicable to the real world studied by the physicists and chemists. On the other hand, the infinite seemed also to be contradicted on the largest possible scale, the universe as a whole. "We need to investigate the expansion of the universe [*Ausdehnung der Welt*], to establish if it results into something infinitely large." Of course, Hilbert's words should not be mistaken for an anticipation of what a few years later became known as the expanding universe. Well aware of the new curved-spaced cosmology based on Einstein's theory of general relativity, he sketched its essence to the mathematicians gathered in Münster [Hilbert 1925, p. 165]:

> The abolition of Euclidean geometry is today not just a purely mathematical or philosophical speculation, but we have also arrived at it from quite a different perspective that originally had nothing to do with the question of the finitude of the world. Einstein has proved the necessity of deviating from Euclidean geometry. Based on his theory of gravitation he attacks the cosmological questions and demonstrates the possibility of a finite world. Moreover, all the results found by the astronomers are also in full agreement with the assumption of the elliptic world.

In the Göttingen lectures he discussed in more detail Einstein's static and finite universe, including the relation between the radius of curvature and the mean density of matter.[3] According to Hilbert, the stars and nebulae were distributed in a huge ellipsoid with diameter about 300,000 light years and thickness about 150,000 light years. Outside the Milky Way, but still within the much thicker ellipsoid, "lie the globular clusters and spiral nebulae which thus have to be counted as parts of this Milky Way system" [Hilbert 2013, p. 710]. His picture of the stellar universe was thus qualitatively similar to the one advocated on the basis of "statistical cosmology" by Cornelius Kapteyn in the Netherlands and Hugo von Seeliger in Germany [Paul 1993; Smith 2006]. According to the "Kapteyn universe" of 1922, the stellar system was about 59,000 light years along the galactic plane and 7,800 light years at right angles to it.

---

[3] Hilbert wrote Einstein's relationship as $2R = \alpha \rho^{-\frac{1}{2}}$, where $R$ is the radius and $\rho$ the mean density of matter in the universe. The constant $\alpha$ relates to Einstein's gravitational constant $\kappa$, or $8\pi G/c^2$, as $\alpha^2 = 2/\kappa$. In a handwritten note he stated that $2R \cong 5 \times 10^9$ light years [Hilbert

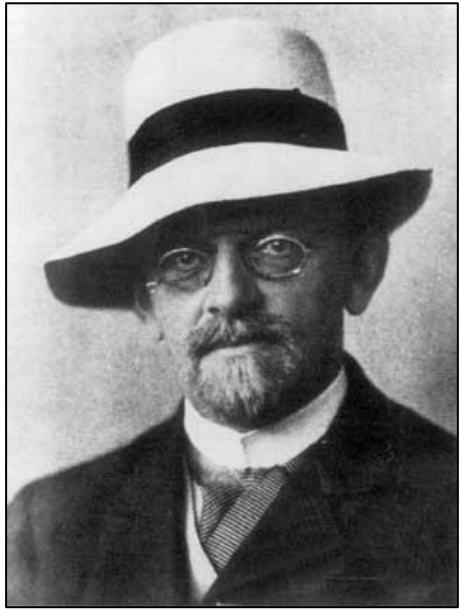

Fig. 2. David Hilbert (1862-1943).

However, the dimensions of the system mentioned by Hilbert were considerably greater. It looked much more like the monster galactic system that Harlow Shapley had suggested in 1918 and which for a period was much discussed. Shapley concluded that the stellar system was approximately 300,000 light years in diameter and more than 30,000 light years in thickness. While Hilbert may have known of Shapley's Milky Way universe, he was apparently unaware of Edwin Hubble's sensational announcement that the distance to the Andromeda was about 930,000 light years and that the spirals were thus "island universes" in their own right. Hubble's discovery was announced on 1 January 1925, but at the time Hilbert gave his lecture it was known only to few astronomers outside the United States. It took several months until it appeared in print [Hubble 1925].

---

2013, p. 709]. Astronomers' estimates of the radius of the Einstein universe were at the time typically about $10^8$ light years [Peruzzi and Realdi 2011].

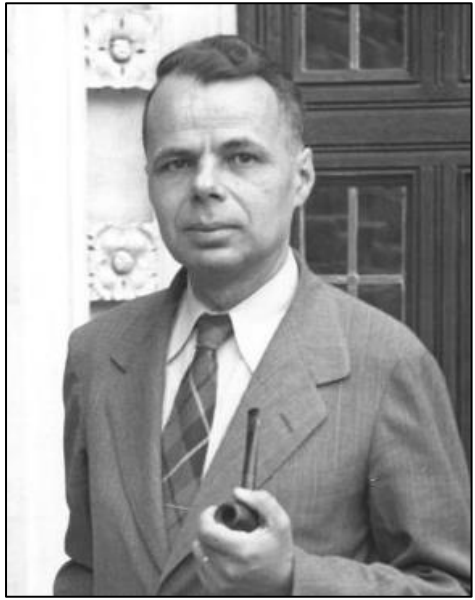

Fig. 5. Lothar Nordheim (1899-1985).

## 3. January 1925: The birth of Hilbert's hotel

The lecture notes for the Göttingen winter semester course were prepared and written up by Lothar Nordheim, a young German physicist who served as one of Hilbert's assistants. In 1923 he obtained his doctoral degree under Max Born on a detailed analysis of the hydrogen molecule according to the rules of the old quantum theory. Nordheim subsequently worked with John von Neumann on aspects of mathematical physics, and in 1928 the two, together with Hilbert, wrote an important paper on the mathematical foundation of quantum mechanics [Hilbert, von Neumann, and Nordheim 1928].[4] He later recalled about his time as an assistant to Hilbert [Wheaton 1977]:

> I was hired by Hilbert to help him in physics. This was in statistical mechanics, which always very much interested me. Then quantum mechanics, the old quantum mechanics. And then when the new quantum mechanics and wave mechanics came out … During the time I was his assistant, he was very sick. And not the genius he had been. He lived very much in the past in a way. His mathematical interest was logic, which was not terribly appealing to me. But he had the conviction that the best thing for a young man was to work with him. That was a reward in itself. And everything else, financial and family considerations, would be way down in importance.

---

[4] Together with Ralph Fowler, Nordheim explained in 1928 the phenomenon of field electron emission in terms of quantum-mechanical tunnelling. As a Jew, he was forced to leave Germany in 1934. He moved to Duke University, USA, where he stayed during most of his active career. During the war years he played an important role in the Manhattan Project.

In his Göttingen lectures Hilbert endeavoured to make clear to his audience the crucial difference between finite and infinite sets. This he did by means of two examples, one related to a hotel and the other to a dance party. If a hotel has only a finite number of rooms, all of them occupied, there is no way to accommodate new guests. For a finite set a part of the set is always smaller than the total set, but this is not the case for an infinite set [Hilbert 2013, p. 730]:

> We now assume that the hotel has infinitely many rooms numbered 1, 2, 3, 4, 5, … and that each of the rooms is occupied by a single guest. All that the manager has to do in order to accommodate a new guest is to make sure that each of the old guests moves to a new room with the number one unit larger. In this way room 1 becomes available for the new guest. One can of course make room for any finite number of new guests in the same manner; and thus, in a world with an infinite number of houses and occupants there will be no homeless.

Hilbert continued:

> The situation is the same with an infinite dance party where all the gentlemen have asked the ladies to dance. A new lady enters, but the organizer of the dance can easily arrange that she will not be without a partner. It is even possible to get space for an infinite number of new guests, respectively ladies [that is: partners for an infinite number of new ladies on the dance floor]. One could, for example, ask the old guest who originally occupied room number n to move to room number 2n. In this way infinitely many rooms with odd numbers would be left free for new guests.

This is what Hilbert had to say about his hotel in January 1924. It was merely an example and one that he attached no particular importance to. Nor did other people at the time find it important. Had the hotel not been resuscitated by Gamow more than two decades later it might well be unknown today. The only allusion to it before 1947 that I know of is from a textbook on calculus published in 1938 and written by Otto Haupt, a mathematician at Erlangen University.[5] Without referring to Hilbert by name, the book posed the following problem: "A hotel with infinitely many rooms $Z_1$, $Z_2$, … is completely occupied. Yet there enters again infinitely many guests $G_1$, $G_2$, … . How is this possible without making one of the existing guests homeless and without more guests having to occupy the same room?"

---

[5] Haupt and Aumann 1938, p. 19. Otto Haupt (1887-1988) served 1921-1953 as professor of mathematics at Erlangen University. For a brief appreciation, see *The Mathematical Intelligencer* **9** (1987): 50-51. His co-author, Georg Aumann (1906-1980), was professor at the University of Frankfurt am Main.

In Göttingen Hilbert used the occasion to extol the virtues of Cantor's set theory, noting that Kronecker had done what he could to fight it. "When I was a *Privatdozent*," he said, "those who believed in Cantor were held in contempt."[6] Even today, he went on, distinguished mathematicians such as Luitzen Brouwer and Hermann Weyl continued resisting Cantor's theory. But, "then we expel from the paradise not only the devil but also all the angels, and then we largely change Cantor's paradise into a wasteland" [Hilbert 2013, p. 742]. While Praising Cantor's theory of the transfinite, Hilbert stressed in his conclusion that the infinite was merely an idea and that it had no role to play whatever in the real world. In fact, the conclusion of his 1925 paper in *Mathematische Annalen*, quoted above, was taken over verbatim from his Göttingen lectures.

## 4. Gamow's hotel

24-year-old George Gamow spent the summer months of 1928 as a postdoc in Göttingen, where he made his breakthrough in theoretical physics by explaining alpha radioactivity on the basis of quantum mechanics [Gamow 1970; Reines 1972]. He most likely met Hilbert and his mathematical colleague Richard Courant, although there is no documentary evidence supporting the conjecture. He also must have met Nordheim, a physicist like himself and Hilbert's former assistant. Many years later, after having migrated to the United States to become a professor at the George Washington University, he pioneered the theory of the early universe that would eventually be known as the big bang theory. His first scientific paper on the subject dates from 1946. Gamow was at the time known also as a successful and original writer of popular science, a reputation based on his classic *Mr. Tompkins in Wonderland* from 1939 and *The Birth and Death of the Sun* from 1940.

In 1947 he followed up on his previous successes with a book titled *One, Two, Three … Infinity* in which he covered in his inimitable style a wide range of problems, ranging from number theory and topology to relativistic space-time and entropy. The book, which in 1961 appeared in a revised edition, also included chapters on "Modern Alchemy," "The Days of Creation" and "The Riddle of Life." It was in this book that he presented Hilbert's hotel as "an example taken from one of the stories about the famous German mathematician David Hilbert" [Gamow 1947, p. 17]. In a footnote he added a reference to "the unpublished, and ever never written, but widely circulating volume: 'The Complete Collection of Hilbert Stories' by R. Courant." Gamow was the eternal prankster and famous for his numerous jokes [Reines 1972]. The reference to Hilbert was very much in the Gamow style, apparently just one more joke. However, it was more than that, for there is little doubt that he was informed about the hotel story during his stay in Göttingen, either by Courant or by Nordheim.

---

[6] After having obtained his doctorate in 1885, Hilbert worked as a *Privatdozent* (private senior lecturer) at the University of Königsberg until he was appointed professor of mathematics in Göttingen in 1895.

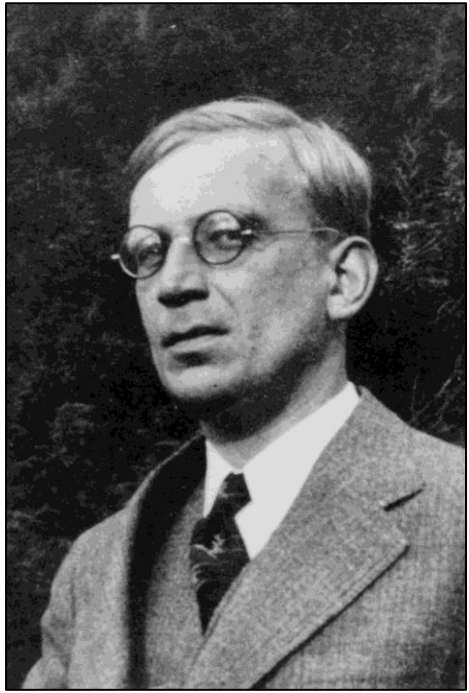

Fig. 4. George Gamow (1904-1968).

Gamow's interest in the riddles of mathematics probably went back to his early university education. As a student at the University of Odessa he initially studied pure mathematics, learning "real mathematics, like the theory of numbers, topology, theory of infinity, and things like this" [Weiner 1968]. His interest was later stimulated by discussions with his friend Stanislaw Ulam, a Polish-born mathematician. Ulam [1972] recalled how they discussed the role of the actual infinite in the physical world and Gamow's belief that the number of objects (from electrons to galaxies) in the universe would be actually countably infinite. According to Ulam [1972, p. 275], "When I acquainted him, sometime in the late '30's, with Godel's [Gödel's] result on undecidability in systems of mathematics, we agreed that this question of the finiteness might be undecidable."

At the time Gamow introduced Hilbert's hotel he was beginning to focus seriously on the structure and evolution of the universe at large. As to its spatial extension, he argued from astronomical measurements that it was infinite. This was his conclusion even before he considered the birth of the universe in a super-dense, primordial state. In a paper written with Edward Teller in 1939 he suggested that the geometry of space was hyperbolic and consequently concluded in favour of "the hypothesis that space is infinite and ever expanding" [Gamow and Teller 1939, p. 657]. He repeated the view in *The Birth and Death of the Sun* [Gamow 1940, p. 202] and maintained it in all his later works on cosmology.

The discussion of Hilbert's hotel in his popular book of 1947 should probably be seen as more than just an intriguing mathematical paradox. It was his way of explaining to readers that the ghost of infinity was not menacing to his favoured cosmological model. The possibility of an infinite universe, described by either a hyperbolic or Euclidean geometry, was not new at the time. As early as 1932 Einstein and the Dutch astronomer Willem de Sitter had jointly proposed a flat model with a beginning in time, and yet they were silent about both the infinity problem and the singularity problem. Gamow may have been the first to call attention to the problem of an actual infinity of objects in open cosmological models. Characteristically, he did it in the context of popular science and in his own way, half serious and half humorous.

In his semi-popular and widely read *The Creation of the Universe* Gamow returned to Hilbert's paradoxical hotel, which he used to convince the reader that there was no contradiction in speaking of an expanding or contracting infinite space. Having recounted Hilbert's alleged story (as he gave it "in one of his lectures"), he commented: "In exactly the same way that an infinite hotel can accommodate an infinite number of customers without being overcrowded, an infinite space can hold any amount of matter and, whether this matter is packed far tighter than herrings in a barrel or spread as thin as butter on a wartime sandwich, there will always be enough space for it" [Gamow 1952, p. 36]. This was the last time that Gamow mentioned Hilbert's hotel. It does not appear in either his autobiography [Gamow 1970] or in the extensive interview that Charles Weiner conducted with him shortly before his death [Weiner 1968].

**5. Hilbert's hotel in philosophy**

For a long time Hilbert's hotel remained unnoticed. The next time it turned up in the literature, as far as I am aware, was after Gamow's death and then without mentioning him. During the 1950s and early 1960s the cosmological scene was marked by a major controversy between evolution models based on general relativity and the new steady state model proposed in 1948 by Fred Hoyle, Hermann Bondi, and Thomas Gold [Kragh 1996]. According to the latter theory, not only did the expanding universe exist eternally with the same average matter density as observed today, it also followed that the curvature constant was zero, implying a spatially as well as temporally infinite universe. Spatial infinity was a possibility within the class of evolution models as well, but in the case of the steady state model it was a necessity. In other words, could it be shown that an infinite number of physical objects were impossible, the steady state model would be impossible as well.

It was in this context that Hilbert's hotel and similar arguments (such as Bertrand Russell's "Tristram Shandy paradox") turned up in the philosophical literature. There were in the early 1960s a couple of attempts based on set theory to show that the infinity of particles in the steady state universe led to contradictory or highly bizarre consequences, but none of them referred explicitly to Hilbert's hotel [North 1965, pp. 379-383; Kragh 1996, pp. 235-236]. In any case, arguments of this kind were scarcely noticed by the involved physicists and astronomers and they did not affect the outcome of the cosmological controversy.

In a paper of 1971 Pamela Huby, a philosopher at Liverpool University, discussed a modernized form of Kant's first cosmological antimony. In a purely philosophical way and without referring to scientific cosmology, she reconsidered the classical question of whether an actual infinity is possible, or whether the universe could be infinite in extension. She argued that this was not the case and that those who still maintained the possibility would have to "accept the many paradoxes that follow … [such as] Hilbert's hotel" [Huby 1971, p. 128]. Huby briefly described the paradoxical hotel, as if it were well known to her readers and without offering a reference to it. Based on logical arguments, among which Hilbert's hotel was one, she concluded that "the universe must be finite in space and … it must have existed for only a finite time."

Whereas Huby was not concerned with the astronomers' universe, in a critical commentary N. W. Boyce [1972; see also Huby 1973] pointed out that Hilbert's hotel described more than just a logically possible world:

> In fact, "Hilbert's Hotel" describes, metaphorically, the structure of the Universe as it is conceived by the "Steady State" Cosmology – that is if the "Steady State" theory of the Universe is true, then, we are living in something very like "Hilbert's Hotel." … The

> "Steady State" theory is exactly like "Hilbert's Hotel" in that an infinite number of guests (galaxies) occupy an infinite number of rooms (space), and in that room is always found for newly arriving guests (galaxies) by moving the existing guests (galaxies) into the rooms (spaces) next door. Thus "Hilbert's Hotel" is no mere mathematical fiction, but, may be the world we actually live in.

This was more or less what Gamow had said in his book of 1952, except that Gamow did not speak of the steady state universe (which he much disliked), but of the infinitely large universe in general. As far as astronomers and physicists were concerned, the steady state theory had been proved wrong by the discovery of the cosmic microwave background in 1965. All the same, philosophers continued to discuss the problem, perhaps unaware that it had already been settled or just considering the solution irrelevant to their logical exercises.

## 6. The theological arena

The discussion of actual infinities has so far been concerned with space or number of particles, but as Huby realized, the problem relates to any kind of actual infinity, whether spatial or temporal. It is in the latter context that it, and Hilbert's hotel as well, becomes relevant to theology and philosophy of religion.

The so-called cosmological proof for the existence of God, dating back to Islamic and Christian philosophers in the early Middle Ages [Sorabji 1983], is based on the claims that (i) the universe has only existed in a finite period of time and (ii) everything in the universe is contingent. Since the universe began to exist, there must be an ultimate cause for its existence and this cause must be a necessary being (an *ens per se*), that is God. The first mentioned claim can be argued empirically, by means of cosmological theory and measurements, or it can be based on logical and mathematical arguments. If it can be proved that an infinite past time is impossible or unreal, it will be a strong argument for the existence of a creative God.

William Lane Craig, an American philosopher and Christian apologist, has been instrumental in reviving the cosmological proof and using set theory in the service of faith. His main argument, a more elaborate version of Huby's reasoning of 1971, is that an infinite temporal regress of events is an actual infinite and that such a concept can have no real existence. While Craig admits that Cantor's set theory is consistent and allows the actual infinite, he considers it a purely mathematical system applicable only to a universe of discourse. When Hilbert's hotel is translated from abstract mathematics to the real spatio-temporal universe, absurdities inevitably follow. In any realistic sense, he concludes, "Hilbert's Hotel is absurd" [Copan and Craig 2004, p. 204].

In a monograph and a paper of 1979 Craig supported his line of argument by referring to Gamow's book of 1947 and its account of the "intellectual creation of the great German mathematician David Hilbert" [Craig 1979a; Craig 1979b]. Not only did he use Hilbert's hotel apologetically, he also extended it by pointing out that it is even weirder if one considers the guests checking out of it. He asked, rhetorically: "Can anyone believe that such a hotel can exist in reality?" While the theist Craig could not, the atheist Gamow could. Ever since Craig's discussion of Hilbert's hotel in the late 1970s it has been a staple topic of conversation in the never-ending discussion of the existence of God as the creator of the universe [Halvorsen and Kragh 2011]. Recently Craig's use of the infinite hotel has been criticized by Landon Hedrick [2014], who argues that the hotel argument does not preclude the possibility of an eternal past.

**7. Conclusion**

The imaginary Hilbert's hotel provides an instructive example of the paradoxes inherent in the notion of the actual infinite. The hotel story was introduced by David Hilbert in lectures that he gave in Göttingen in January 1925, when he also illustrated the difference between finite and infinite countable sets with an example of a dance party. He did not refer to the infinite hotel in his writings, and it only turned up in print in a book Gamow published in 1947. Whereas Hilbert used the story to point out that the actual infinite cannot be part of reality, to Gamow it served as an illustration of the spatially and materially infinite universe. A few philosophers have used Hilbert's hotel to examine conceptually the infinitely large steady state universe and since the late 1970s it has frequently appeared in theological discussions concerning the age and size of the universe. Today the hotel has grown into a minor industry.

**Acknowledgements**: I am grateful to Tilman Sauer, Caltech, who pointed out to me that Hilbert's hotel is in fact due to Hilbert and not, as I first thought, to Gamow. I also want to thank Klaus Scharnhorst, Vrije Universiteit Amsterdam, for references to early German literature that mentions the infinite hotel.